\documentstyle[prl,aps,epsf]{revtex}
\draft
\begin{document}
\twocolumn[\hsize\textwidth\columnwidth\hsize\csname @twocolumnfalse\endcsname

\title{Vortex Pinball Under Crossed AC Drives in Superconductors
with Periodic Pinning Arrays}
\author{C. Reichhardt and C.J. Olson} 
\address{ 
Center for Nonlinear Studies and Applied Theoretical and
Computational Physics Division, 
Los Alamos National Laboratory, Los Alamos, NM 87545}

\date{\today}
\maketitle
\begin{abstract}
Vortices  
driven with both a transverse and a longitudinal AC drive
which are out of phase are shown to exhibit a novel 
commensuration-incommensuration effect when interacting with 
periodic substrates.      
For different AC driving parameters, the motion of the vortices 
forms commensurate orbits with the periodicity of the pinning array. 
When the commensurate orbits are present,
there is a finite DC critical depinning threshold, while for the
incommensurate
phases the vortices are delocalized and the
DC depinning threshold is absent.  
\end{abstract}
\vspace{-0.1in}
\vspace{-0.3in}

\vskip2pc]
\narrowtext

A wide variety of dynamical systems   
can be modeled as a classical
particle moving in a periodic potential.
Examples of such systems
are Josephson-junction (JJ) arrays \cite{Shapiro,Benz}, 
sliding charge density waves (CDW) \cite{Gruner,Thorne}, 
atomic friction \cite{Persson}, vortices
in superconductors moving over periodic substrates 
\cite{Martinoli,Moshchalkov,Reichhardt,Look,Zimanyi,Martin,Kolton}, 
electrons at 
low magnetic fields in 
antidot arrays \cite{Weiss,Geisel}, 
as well as ions \cite{Petsas} and colloids
\cite{Korda} in 
optical trap arrays.
When an AC drive is superimposed over the DC drive,
resonances or phase-locking effects can occur which have been 
extensively studied in JJ arrays \cite{Benz} and CDWs \cite{Gruner}. 
Signatures of the phase locking include oscillations in the
depinning threshold with increasing AC drive amplitude,
and locking of the particle velocity 
to a fixed value over a range of the DC drive. 
In superconductors
with periodic substrates, phase locking of driven vortices has been
experimentally observed in
samples containing 1D periodic modulations \cite{Martinoli}. 
More recently, experiments, simulations, and theory
have shown that in superconductors with periodic hole arrays,
phase locking or Shapiro steps occur when combined 
DC and AC drives are applied.

In all these phase locking systems the AC and DC drives are in 
the {\it same direction} and the motion can be 
considered effectively one-dimensional (1D). 
Different effects can arise in a two-dimensional (2D) system.
Recently it was shown that 
new kinds of phase-locking effects occur when the AC drive is 
applied {\it perpendicular} to the DC drive for systems with periodic
substrates \cite{Kolton}. 

Here we study vortex motion in superconductors with periodic arrays of
pinning sites and {\it three} combined driving forces.
A DC drive $f_{dc}$ is applied along the longitudinal direction,
and {\it two} AC drives which are 90 degrees out of phase are 
added
in the longitudinal
and transverse directions. 
In the absence of pinning and at $f_{dc}=0$, the vortices move in a
circular orbit with radius and eccentricity 
determined by the amplitudes and
frequencies of the AC drives. 
We focus on a regime just above the first matching field $B_{\phi}$,
where each pin is occupied by one vortex and a small
number of additional vortices move in the periodic potential created 
by the pinned vortices.
The existence of such interstitial vortices above $B_{\phi}$
has been inferred from transport measurements and direct
imaging in samples with periodic holes \cite{Moshchalkov},
and in this same regime
Shapiro steps have previously been observed 
\cite{Look}.

The system we propose has 
similarities to the ``electron pinball'' model studied by 
Weiss {\it et al.} \cite{Weiss} and others \cite{Geisel}, in which 
classical cyclotron electron motion is induced by a 
magnetic field in samples with periodic arrays of anti-dots. 
In these systems, peaks and dips appear in the magnetoresistance
as a function of field.  The features are 
believed to arise when classically moving electrons
follow {\it pinned} or {\it commensurate} circular orbits
enclosing integer numbers of dots, in which
the electrons can travel
without scattering off the dot potentials.  
At the {\it incommensurate} orbits the electrons are 
scattered and diffuse throughout the sample.   
There are important differences between the vortex 
"pinball" system 
and the electron systems. 
The vortices interact with a  {\it long-range} smooth
or egg-carton square potential, rather than simple point scatters,
and thus
the moving vortices
must follow square orbits (in the case of square pinning arrays), 
rather than circular orbits. 
A particularly attractive feature of the vortex system is that the 
{\it shape} of the orbit can be 
carefully controlled experimentally
by changing the ratio of the AC amplitudes, phases or
frequencies. This allows new kinds of 
anisotropic orbits and commensuration effects to be produced,
which are not obtainable in the electron pinball models.  
 
We consider a thin superconductor containing 2D vortices,
which is the appropriate model for
the recent experiments in superconductors 
with hole and dot arrays.
The vortex-vortex interaction has the form of a
logarithmic potential, $U_{v} = -A_{v}\ln(r)$, with the energy normalization
$A_{v} = \Phi^{2}_{0}/8\pi \Lambda$. Here, $\Phi_{0}$ is the flux
quantum and $\Lambda$ is the effective 2D penetration depth for 
a thin film superconductor.   
The overdamped normalized 
equation of motion for a single vortex $i$ is
\begin{equation}
{\bf f}_{i} = \eta\frac{d  {\bf r}_{i}}{dt} = 
{\bf f}_{i}^{vv} + {\bf f}_{i}^{vp} + {\bf f}_{dc} + {\bf f}_{ac} = 
\eta{\bf v}_{i}
\end{equation} 
where the damping term $\eta$ is the Bardeen-Stephen friction.
The force from vortex $i$ on the other vortices is

\begin{figure}
\center{
\epsfxsize=3.5in
\epsfbox{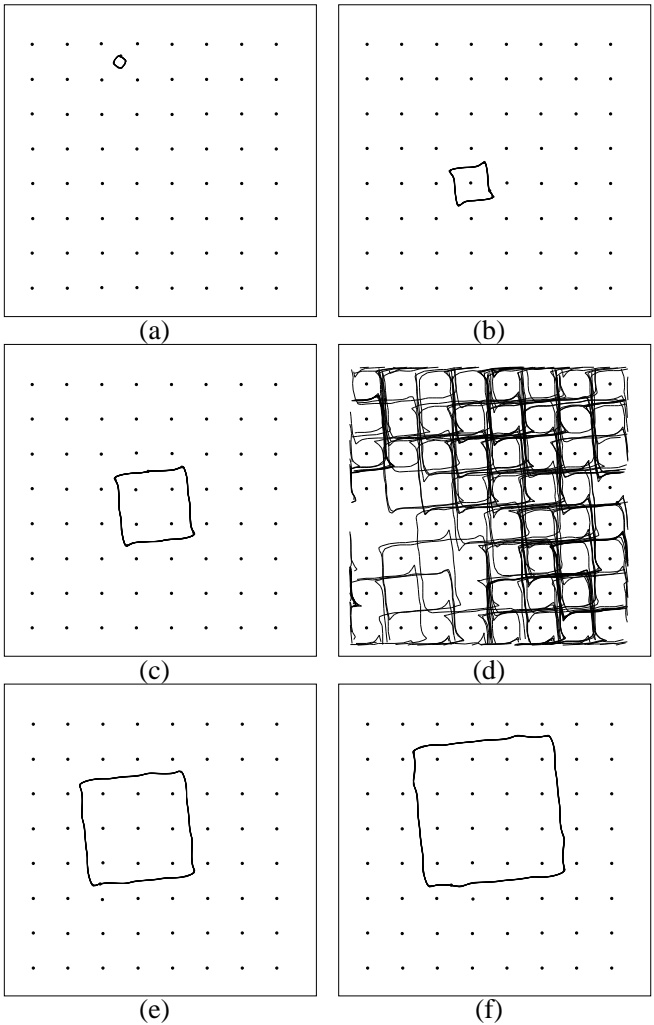}}
\caption{
Vortex positions (dots) and trajectories (lines) 
for $f_{dc} = 0.0$ and different isotropic AC amplitudes.
(a) $A = 0.16$, (b) $A = 0.21$, (c) $A = 0.3$, (d) $A = 0.355$,
(e) $A = 0.39$, and (f) $A = 0.47$. 
}
\end{figure}

\hspace{-13pt}
${\bf f}_{i}^{vv} = -\sum_{j\neq i}^{N_{v}}\nabla_{i} U_{v}(r_{ij})$.
The long range 
interaction is treated
with a fast converging summation method 
\cite{Gronbech}. 
The pinning force ${\bf f}^{vp}$ arises
from the pinning sites which are modeled as short
range attractive parabolic wells. The pinning sites are placed in a square
array of side $L$, and each pin has a radius of 
$r_{p} = 0.15L$, which
is within the typical experimental ratios of $r_{p}/L = 0.14$ to $0.3$.   
The DC driving term ${\bf f}_{dc}$ is applied
along the symmetry axis of the  
pinning array, in the $x$-direction. 
The
AC driving term is ${\bf f}_{AC} = A\sin(\omega_{A}t){\hat {\bf x}} + 
B\cos(\omega_{B}t){\hat {\bf y}}$, where 
we fix $w_{A}/w_{B} = 1.0$. 
For most of the results presented here we consider the case of a
system containing 64 pins and a
vortex filling fraction
of $B/B_{\phi}=1 + 1/64$;
however, we have found the same results for higher filling
fractions (such as $B/B_{\phi} = 2$, $7/4$, $3/2$, and $5/4$)
at which the
interstitial vortices form a symmetric pattern so that interstitial vortex
interactions cancel.  
The initial vortex position is found by annealing from a high temperature
with no driving and cooling to $T=0$. 
The DC drive is increased in increments of $0.0001$ and the 
sample is held at each drive for $3 \times 10^5$ time steps
to ensure a steady state; 
the DC depinning threshold is determined from
the time averaged vortex velocities $<V_{x}>$.   

We first consider the case of isotropic AC driving with  
$A/B = 1.0$ and 
show that 
commensurate-incommensurate 
depinning 
transitions occur as a function of increasing $A$
even at $f_{dc}=0$.
Example vortex trajectories are
illustrated in Fig.~1.
For $A = 0.16$ [Fig.~1(a)] the vortex is confined to move in a 
small circular orbit in the middle of the plaquette. For $A = 0.21$
[Fig.~1(b)] the AC amplitude is large enough that the 
vortex encircles $n=1$ pin in an orbit that 
is slightly tilted due to the
counter-clockwise vortex movement through
the square potential.
In Fig.~1(c), for $A = 0.3$, the orbit encircles $n=4$ pins. 
In Fig.~1(d),
for $A = 0.355$, the orbit radius is somewhere between 
$n=4$ and $n=9$, and
the vortex interacts strongly with the occupied
pins, scattering off them. 
The vortex becomes delocalized, and diffuses throughout the sample, 
avoiding areas near pins due to the repulsion
from the pinned vortices.
The orbit switches intermittently between $n=4$ and $n=9$.
In Fig.~1(e,f) we show localized vortex orbits for
$A = 0.39$ and $0.47$, 
with $n=9$ and $n=16$, respectively. 
We also find a stable orbit with $n=25$ for $A = 0.57$.
For $ 0.42 < A  < 0.45$, as well as for $0.52 < A < 0.55$, the vortex 
is delocalized and moves in a manner similar to that shown in Fig.~1(d).   

In general
we find localized orbits at values of $A$ for which the vortex
trajectory encloses $n=m^2$ pins, where $m$ is an integer. 
By comparison, in
electrons in antidot lattices, Weiss {\it et al.} \cite{Weiss}
found commensurate or pinned orbits 
when the number of dots encircled was 
$n = 1$, 2, 4, 9, 16, and $21$. We do not observe any stable orbits for
$n = 2$ and $21$. In the electron systems \cite{Weiss,Geisel} 
these orbits correspond to states where
a portion of the electron orbit closely approaches the
dots, which are point scatters. 
In the vortex system, due to the long range
vortex-vortex interactions, 
the mobile vortices cannot follow orbits that approach arbitrarily
closely to the pinning sites.
  
In Fig.~2 we show the
DC depinning threshold $F_{dp}/F^{0}_{dp}$
vs $A$
for the isotropic case $A=B$. 
Here, $F^{0}_{dp}$ is the
DC depinning force for $A=0$.
The commensurate orbits are
{\it pinned} at low $f_{dc}$, producing a series of peaks 
in $F_{dp}$ at the commensurate AC values.
At the incommensurate phases the depinning threshold vanishes. 
For $0.0 < A < 0.2 $, the vortices move in the $n=0$
interstitial orbit illustrated in Fig.~1(a). 
$F_{dc}$ decreases as the 
AC amplitude increases until reaching a minimum value
at $A=0.19$, followed by a peak in $F_{dc}$
for $A = 0.225$, which corresponds to an $n=1$ orbit
[Fig.~1(b)]. The second peak at $A = 0.31$ corresponds to 
the $n=4$ orbit
[Fig.~1(c)].
For the delocalized incommensurate orbits [Fig.~1(d)] of 
$0.34 < A < 0.36$, $F_{dp}$ vanishes. 
A finite $F_{dp}$ 

\begin{figure}
\center{
\epsfxsize=3.5in
\epsfbox{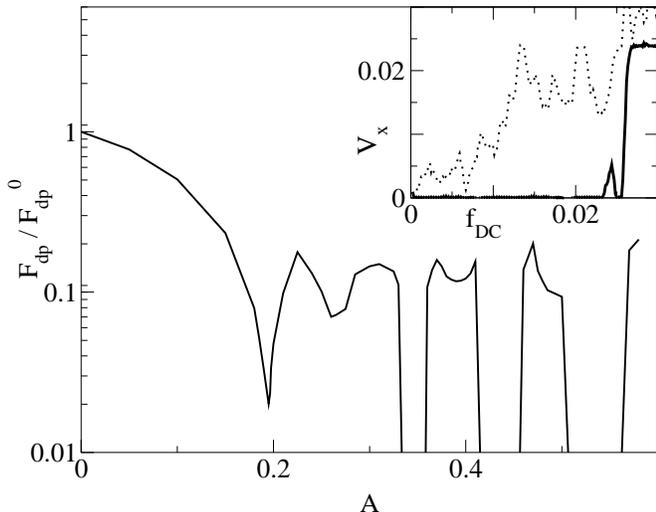}}
\caption{
The DC depinning force $F_{dp}/F_{dp}^{0}$
vs AC amplitude $A$ for the isotropic case $A=B$. 
$F_{dp}^{0}$ is the
depinning force for zero AC drive. 
Inset: average DC velocity $<V_{x}>$ vs DC driving force $f_{dc}$
for an incommensurate orbit $A = 0.43$ (dotted line) and a commensurate
orbit $A = 0.37$ (solid line).   
}
\end{figure}

\hspace{-13pt}
is regained when $n=9$ orbits 
appear for $0.36 < A < 0.42$.
As $A$ continues to increase, $F_{dp}$ is nonzero for $A$ values
at which the stable $n=16$ and $n=25$ orbits are observed, 
with $F_{dp}=0$ in portions of the regions between these values 
of $A$. 
We have checked the oscillatory behavior for different values of 
$L$ and find the same general behavior.  
Since the radius of the vortex orbit 
$R \sim A/\omega_{A}$, 
a similar series of peaks in $F_{dc}$ 
occurs if $A$ is fixed and $1/\omega_A$ is varied.  
In the inset of Fig.~2 we show
typical curves of the average DC 
velocity $<V_{x}>$ vs $f_{dc}$ for an unpinnned incommensurate orbit 
at $A = 0.43$ and a pinned commensurate orbit at $A = 0.37$,
illustrating 
the well defined {\it sharp} depinning threshold for the pinned orbits.  

The maximum $F_{dp}$ values
in Fig.~2 do {\it not} decrease with
increasing $A$, indicating that all of the commensurate orbits are
pinned equally well. 
In addition the boundaries between the pinned and
unpinned regions are sharp.  
Thus the behavior of $F_{dp}$ vs $A$
clearly differs from that associated with Shapiro steps 
\cite{Shapiro,Benz,Zimanyi}, 
where the depinning threshold oscillates 
with $A$ according to
a Bessel function $J_{0}(A)$, the peaks in the depinning
force are smooth, and the peak $F_{dp}$ value
gradually decreases with $A$.

We next consider anisotropic or elliptical orbits for $A \neq B$.
In Fig.~3 we show 
$F_{dp}/F_{dp}^{*}$ vs $B/A$ for  
fixed $A = 0.225$ and varying $B$, where $F_{dp}^{*}$ is the
depinning force for $B = 0$. 
Again we find strongly pinned orbits, indicated by
peaks in $F_{dp}$; 
however, in this case the pinned orbits occur when
the vortex orbit encircles $n=m$ pins. 
If a higher value of $A$ is chosen such that 
the orbit encircles 
two pinning sites in the transverse direction in one period,
commensurate orbits that encircle $n=2m$ pins appear.
The peaks in the depinning curve shown here are much 
more symmetric than those in
Fig.~2. This may be due 

\begin{figure}
\center{
\epsfxsize=3.5in
\epsfbox{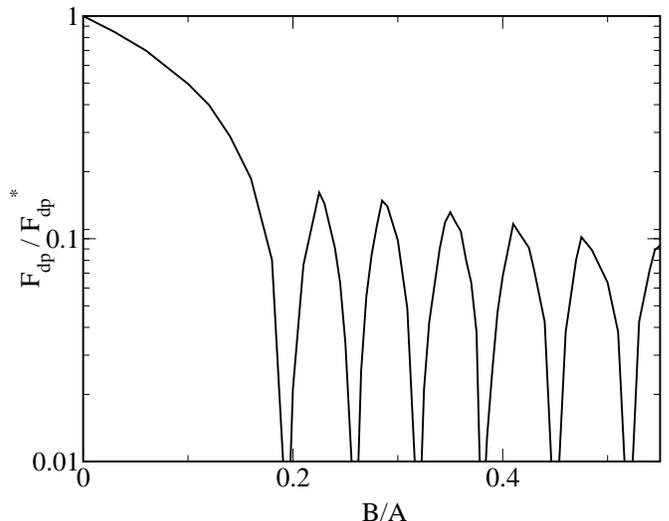}}
\caption{
The DC depinning force $F_{dp}/F_{dp}^{*}$
vs AC amplitude $B/A$ for the anisotropic case
where $A = 0.225$ and $B$ is varied. $F_{dp}^{*}$ is the 
depinning force for $B = 0.0$. 
}
\end{figure}

\hspace{-13pt}
to the fact that the elliptical orbits are
less perturbed by the square pinning potential
than the isotropic circular orbits. 

In Fig.~4 we present several anisotropic commensurate orbits
at the depinning peaks shown in Fig.~3. 
Here the orbits for Fig.~4(a,b,d)
encircle $n=1$, 2, and 3 pins, respectively.
This trend continues for the higher peaks. 
In Fig.~4(c) we also show a {\it sliding} orbit at
$B = 0.3$ with $f_{dc}$ above depinning. 

We now discuss experimental systems in which these phases
can be observed.
For superconductors with periodic antidot arrays, the pin geometry
should be chosen such that the pins have vortex saturation numbers of one, 
so that above $B_{\phi}$ the additional vortices will sit in the
interstitial regions.
For pins with higher saturation numbers,
commensuration effects should still be observable at higher 
matching fields when interstitial vortices start to appear. 
Samples with very low intrinsic pinning should be used to 
enhance the effect.
Samples in which Shapiro steps for moving interstitial vortices
have been observed would be ideal. 
As in the case of Shapiro steps \cite{Zimanyi},
the variation of the depinning force versus AC drive amplitude
should be most pronounced at
filling fractions where the interstitial vortices form a symmetrical 
pattern and the interstitial vortex interactions effectively cancel,   
such as at $B/B_{\phi} =$ 2.0, 1.75, 1.5, and 1.25. 
Using samples such those in Ref.\cite{Look}
where Shapiro steps are observed, 
for a pinning lattice spacing of $a = 2 \mu m$, 
$\omega = 2\pi \nu$, and $\nu = 40 MHz$,
commensuration effects
should be observable with applied crossed AC currents from 
0 to 10 $I_{c}$ where $I_{c}$ is the critical current.   
Superconductors with rectangular and triangular 
(rather than square) pinning arrays should
also exhibit these phenomena; however, 
the stable orbits would encircle different numbers of
pins than the orbits described here. 
In superconductors with 

\begin{figure}
\center{
\epsfxsize=3.5in
\epsfbox{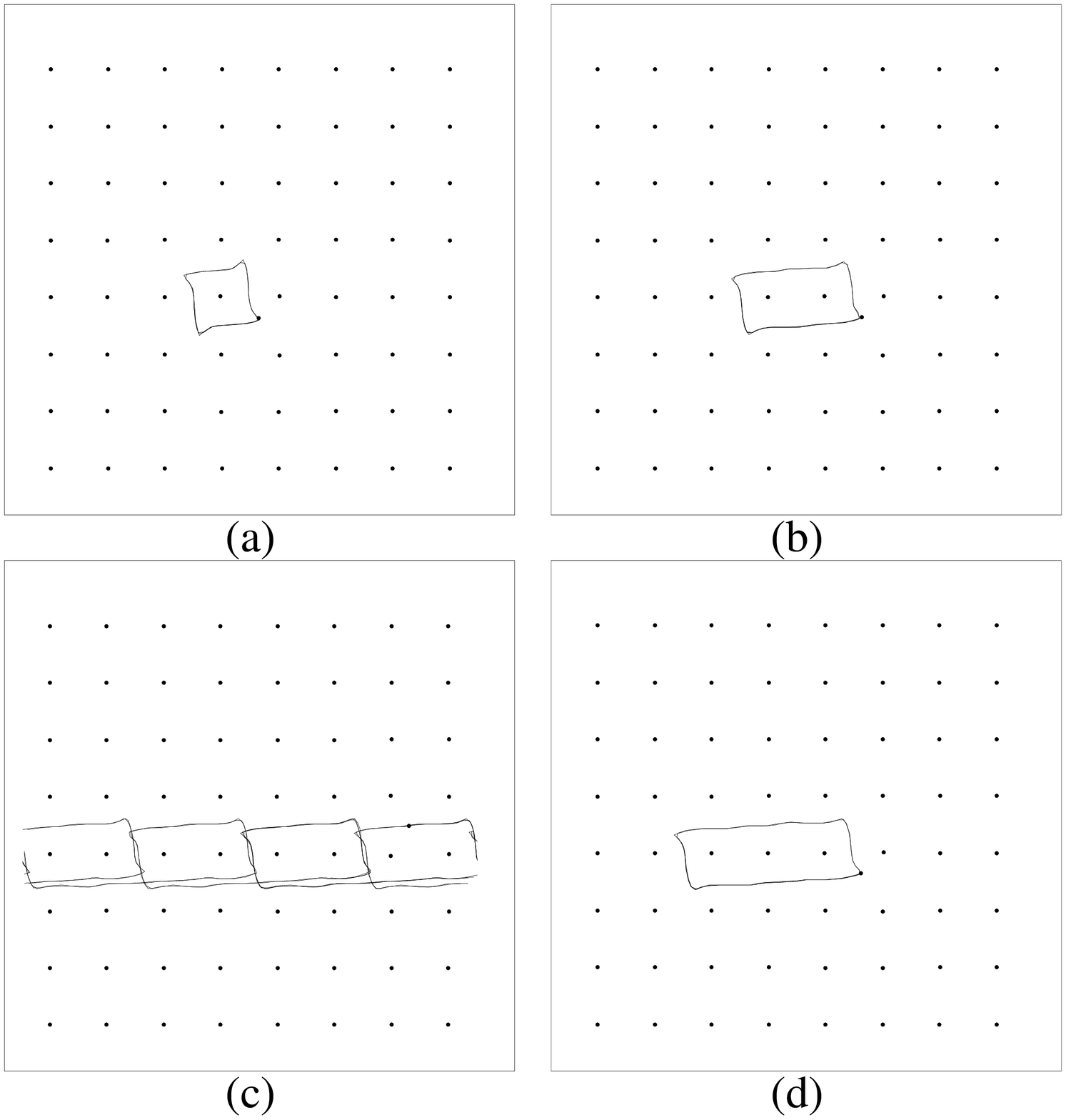}}
\caption{
Vortex positions (dots) and trajectories (lines) for fixed
$A = 0.225$ and (a) $B = 0.225$, (b) $B = 0.285$ and (d) $B = 0.35$.
In (c) a sliding orbit is shown for $B = 0.3$ and $f_{dc} > F_{dp}$. 
}
\end{figure}

\hspace{-13pt}
magnetic dot arrays, the dots can act as 
long range repulsive sites, so that 
pinned orbits of the type described here
could be observed at vortex filling fractions such 
as $B/B_{\phi}=1$ and $1/2$. 
In addition, superconductors with 2D smoothly modulated surfaces
should also produce commensurate and incommensurate orbits. 
Other promising systems 
in which these phases could be observed include 2D JJ
arrays, 2D atomic friction models, and colloids on
optical pin-scapes \cite{Korda}.
A possible application of our results would be 
particle segregation in multi-species
systems, such as colloids on periodic substrates, 
where the different species have 
different mobilities.
Here it should be possible 
to tune the AC drive such that one species would 
be pinned and another depinned.
The species could then be segregated
with a DC drive.

To summarize,
we have numerically studied the motion of vortices 
interacting with a 2D 
periodic potential created by immobile vortices located at 
pinning sites placed in a square array.  We apply
{\it two} AC drives,
perpendicular to one another and out of phase by $90^{\circ}$, 
such that the vortices move in a circle in the absence 
of a substrate potential. 
For AC drives of equal amplitude,
as a function of increasing AC amplitude 
we find a series of 
pinned orbits enclosing $n=m^2$ pinning sites.  Each of these orbits
has a finite depinning threshold to an additional applied DC force. 
At AC amplitude values
between these pinned orbits, the vortices are delocalized
and diffuse through the sample, and the DC depinning threshold is zero.  
Experimentally these states can be observed as a series of pinned and
non-pinned regions as a function of AC amplitude or frequency. 
For anisotropic AC drives,
we find a series of asymmetric pinned orbits 
which enclose $n=m$ pinning sites. 
We call our model the vortex pinball model in analogy to the
electron pinball system for electron cyclotron motion in anti-dot arrays. 
We also suggest other systems in which these
phases can be observed experimentally. 

Acknowledgements: This work was supported by the US Dept. of Energy
under contract W-7405-ENG-36.

\end{document}